\documentstyle[epsfig,graphicx]{aipproc}

\begin{document}
\title{Inertial control of the VIRGO Superattenuator\footnote{To appear in the
Proceedings of the {\it Third E.Amaldi Conference on Gravitational
Waves Experiments}, Caltech, Pasadena, 12-16 July 1999.}}

\author{Giovanni Losurdo$^*$ for Pisa and Florence VIRGO Groups}
\address{$^*$Istituto Nazionale di Fisica Nucleare - Sezione di Pisa\\
Via Livornese 1291 - 56010 - S.Piero a Grado (Pisa) - Italy\\
e-mail: losurdo@galileo.pi.infn.it}

\maketitle

\begin{abstract}
The VIRGO superattenuator (SA) is effective in depressing the
seismic noise below the thermal noise level above 4 Hz. On the
other hand, the residual mirror motion associated to the SA normal
modes can saturate the dynamics of the interferometer locking
system. This motion is reduced implementing a wideband (DC-5 Hz)
multidimensional control (the so called {\it inertial damping})
which makes use of both accelerometers and position sensors and of
a DSP system. Feedback forces are exerted by coil-magnet actuators
on the top of the inverted pendulum. The inertial damping is
successful in reducing the mirror motion within the requirements.
The results are presented.
\end{abstract}

\section{Introduction}

The test mass suspension of the VIRGO detector, the
superattenuator (SA) \cite{sa}, has been designed in order to
suppress the seismic noise below the thermal noise level above 4
Hz. The expected residual motion of the mirror is $\sim 10^{-18}$
m$\sqrt{{\rm Hz}}$ @4 Hz. At lower frequencies, the residual
motion of the mirror is much larger ($\sim 0.1$ mm RMS), due to
the normal modes of the SA (the resonant frequencies of the system
are in the range 0.04-2 Hz).

To lock the VIRGO interferometer the RMS motion of the suspended
mirrors must not exceed $10^{-12}$ m (to avoid the saturation of
the read-out electronics). VIRGO locking strategy is based on a
hierarchical control: feedback forces can be exerted on 3 points
of the SA (inverted pendulum (IP) \cite{ip}, {\it marionetta} and
mirror). The control on the 3 points is operated in different
ranges of frequency and amplitude. The maximum mirror displacement
that can be controlled from the marionetta without injecting noise
in the detection band is $\sim 10$ $\mu$m. Therefore, a damping of
the SA normal modes is required for a correct operation of the
locking system. An active control of the SA normal modes, using
sensors and actuators on top of the IP, capable of reducing the
mirror residual motion within a few microns, has been successfully
implemented.

\section{Experimental setup}

\begin{figure}[t]\begin{center}
\begin{minipage}{5.5cm}
    \includegraphics[bb=2.5cm 1.5cm 17cm
28cm,clip=true,width=6cm]{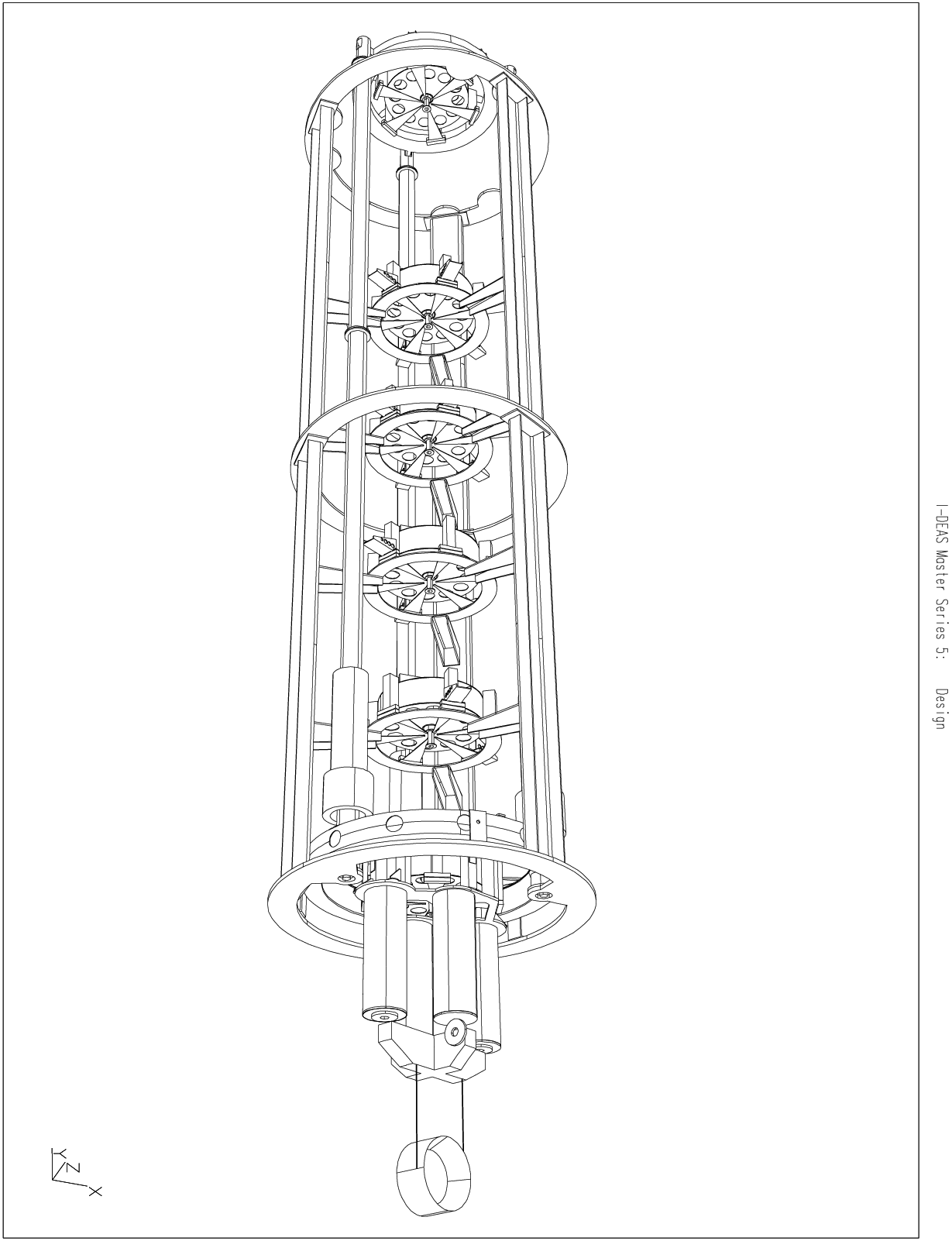}
\end{minipage}\hfill
\begin{minipage}{8.5cm}
 \includegraphics[width=7.5cm]{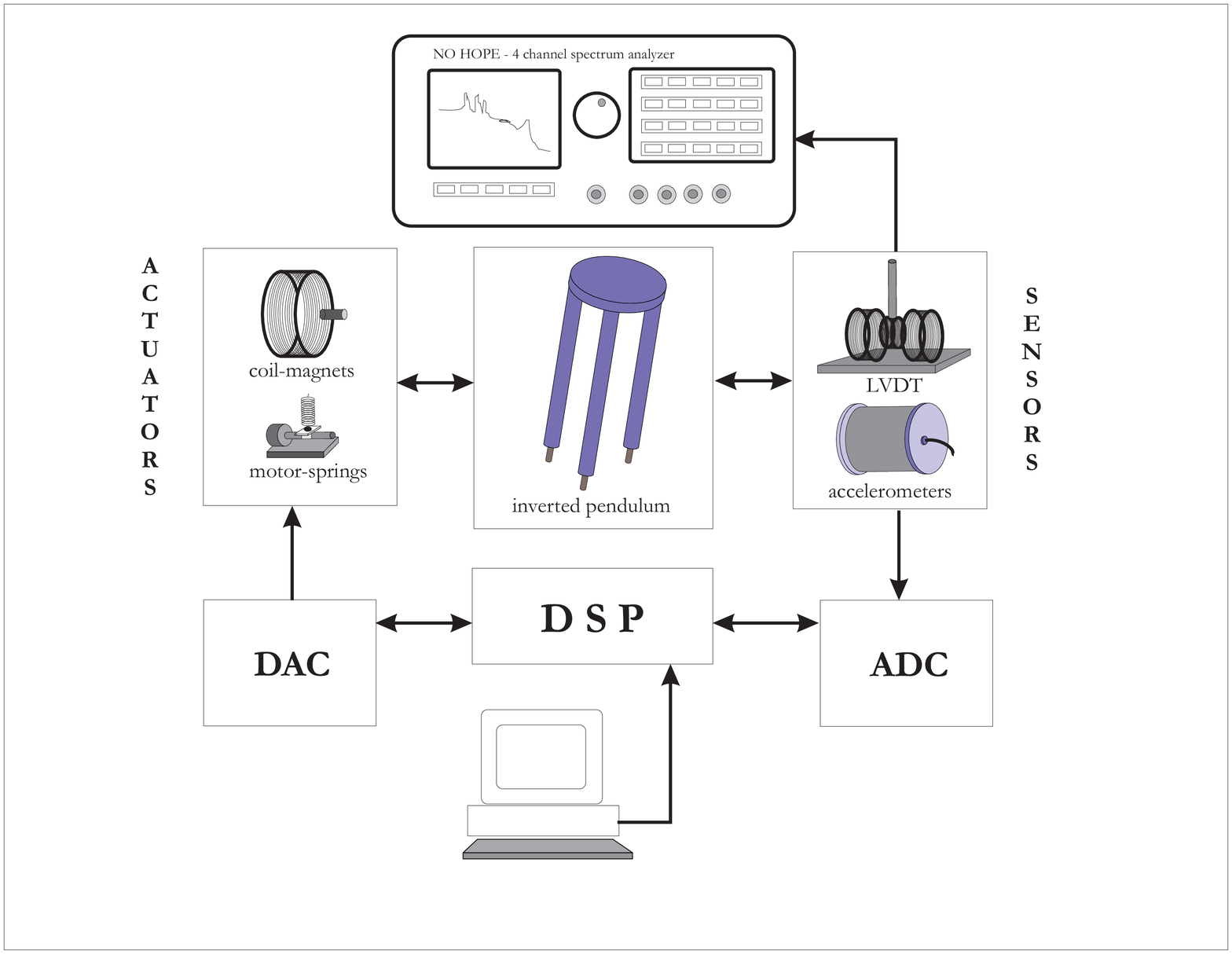}
 \includegraphics[width=8cm]{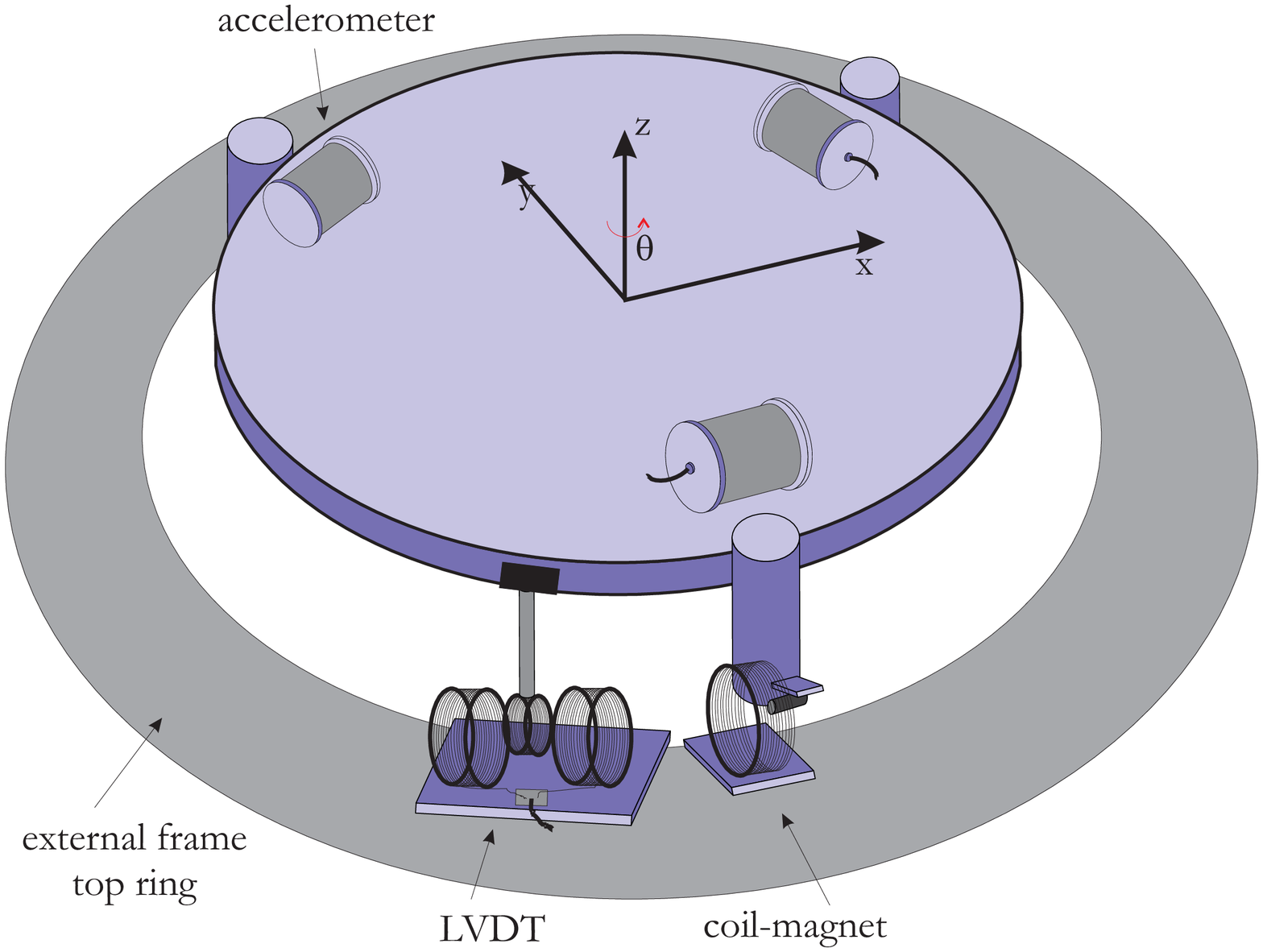}
\end{minipage}
\caption{\footnotesize LEFT: the superattenuator; RIGHT TOP:
logical scheme of the setup for the local active control; RIGHT
BOTTOM: simplified view of the IP top table, provided with the 3
accelerometers. One LVDT position sensors and one coil-magnet
actuator are also shown. } \label{setup} \end{center}\end{figure}

The setup (fig. \ref{setup}) of the experiment is composed by a
full scale superattenuator, provided with 3 accelerometers (placed
on the top of the IP), 3 LVDT position sensors (measuring the
relative motion of the IP with respect to an external frame), 3
coil-magnet actuators. The accelerometers work in the range DC-400
Hz and have acceleration spectral sensitivity $\sim 10^{-9}\; {\rm
m\,s}^{-2}\,{\rm Hz}^{-1/2}$ below 3 Hz \cite{acc}. The sensors
and actuators are all placed in {\it pin-wheel} configuration. The
sensors and actuators signals are elaborated by a computer
controlled ADC (16 bit)-DSP-DAC (20 bit) system. The DSP allows to
handle the signals of all the sensors and actuators, to recombine
them by means of matrices, to create complex feedback filters
(like the one of fig. \ref{filter}) with high precision
poles/zeroes placement and to perform a large amount of
calculations at high sampling rate (10 kHz). The suspended mirror
is provided with an LVDT to measure its displacement with respect
to ground. 

\section{The control strategy}

The active control of the SA normal modes is defined {\it inertial
damping}, because it makes use of inertial sensors
(accelerometers) to sense the SA motion. The advantage of using
accelerometers is that they perform the measurement with respect
to the ``fixed stars", while position sensors do it with respect
to a reference frame which is not seismic noise free. Therefore,
inertial sensors are to be used so that no seismic noise is
reinjected by the feedback. Actually, in the real SA control both
sensors are used: position sensors provide a low frequency (DC -
10 mHz) control of the SA position (in order to avoid drifts),
while accelerometers allow a wideband reduction of the noise in
the region of the SA resonances (10 mHz - 2 Hz).

\begin{figure}[b]\begin{center}
\begin{minipage}{7cm}
\includegraphics[width=7cm]{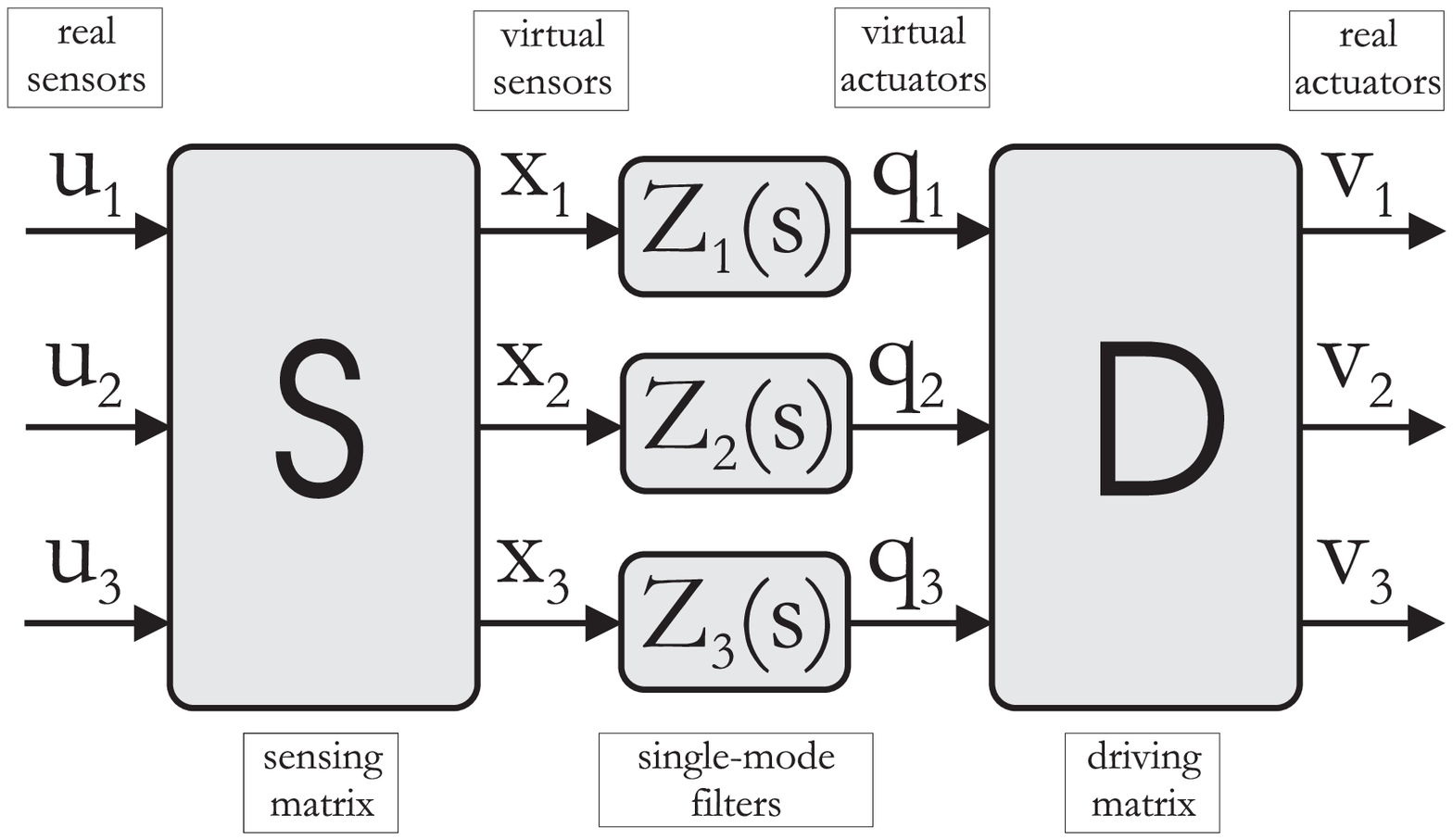}
\end{minipage}
\begin{minipage}{7cm}
\includegraphics[width=7cm]{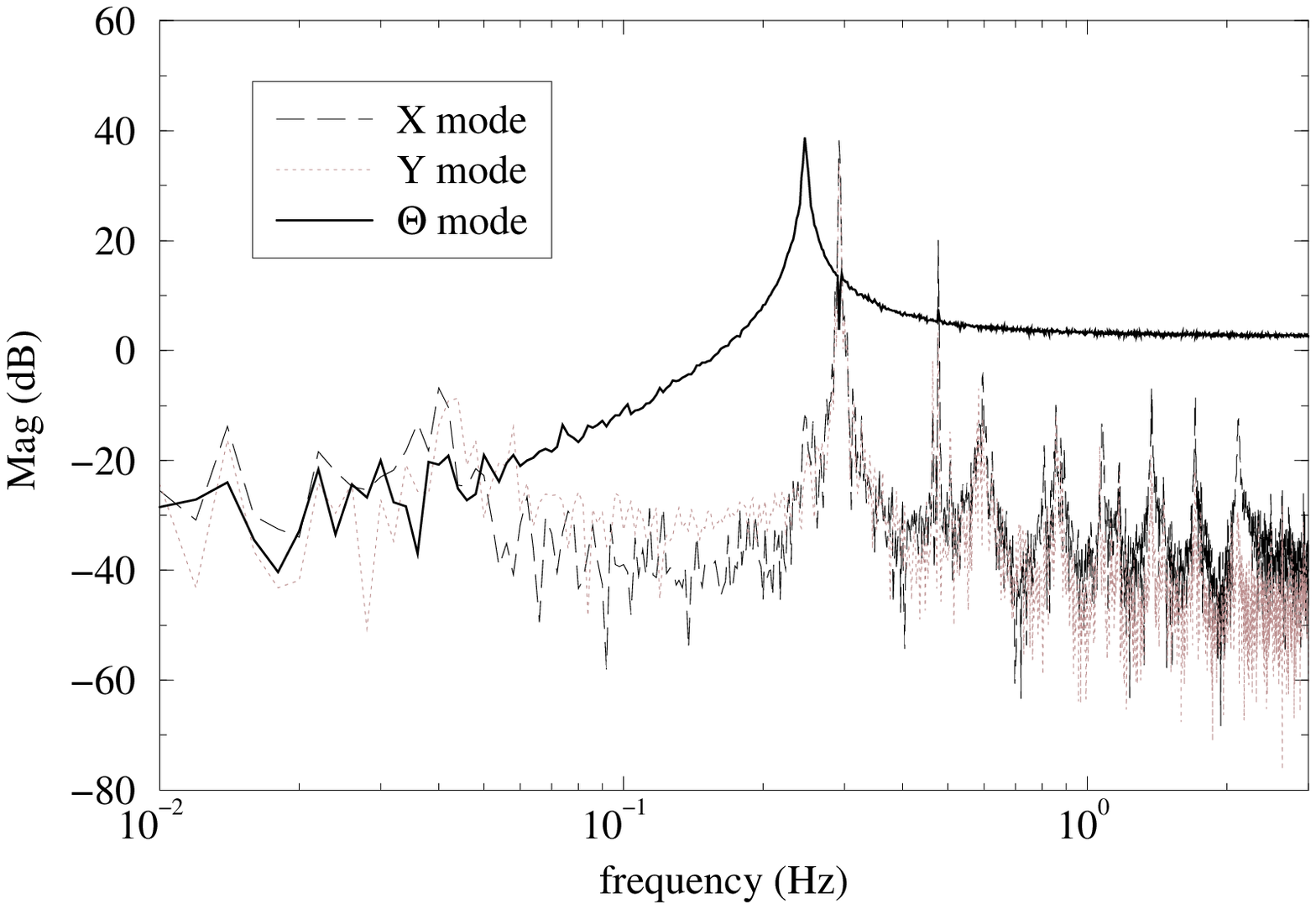} \end{minipage} \caption{\footnotesize
LEFT: The logic of the diagonalisation: the output $u_i$ of the
sensors are linearly recombined by a matrix {\bf S} in order to
produce 3 {\it virtual} sensors outputs ($x_i$), sensitive to pure
modes. Three independent feedback filters $Z_i(s)$ are designed
for the pure modes and 3 generalized forces $q_i$ are produced.
The $q_i$ are turned into real currents ($v_i$) to feed the
actuators via the matrix {\bf D}. RIGHT: Effect of the digital
diagonalization: the 3 modes are uncoupled, the $\Theta$ mode is
excited. } \label{decomp} \end{center}\end{figure}

\begin{figure}\begin{center}
\begin{minipage}{7cm}
    \includegraphics[width=7cm]{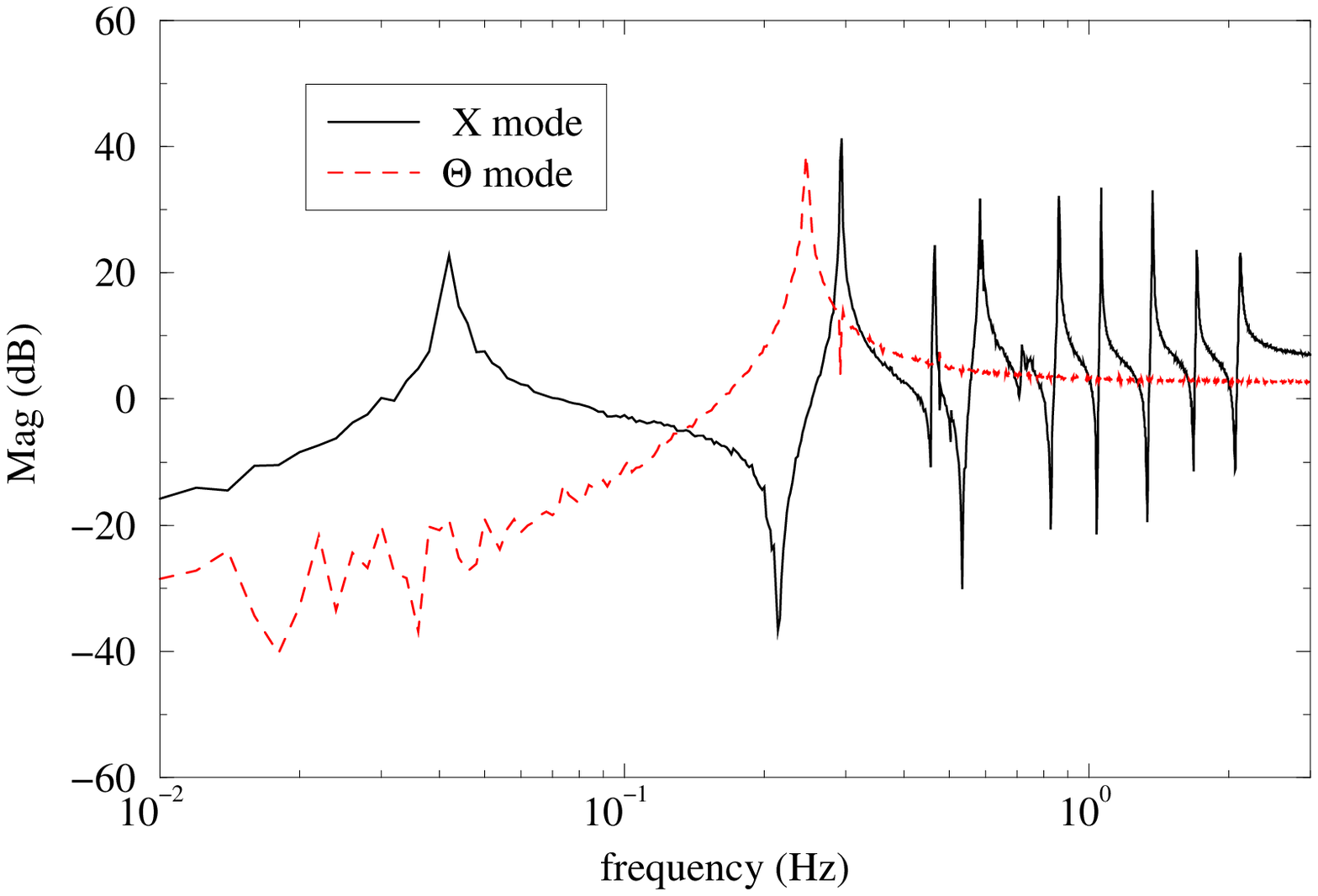}
    \caption{\footnotesize The output of the virtual accelerometers
    $X$ and $\Theta$ are compared. Different feedback strategies are
    required for the two modes} \label{xth}
\end{minipage}\hfill
\begin{minipage}{7cm}
 \includegraphics[width=7cm]{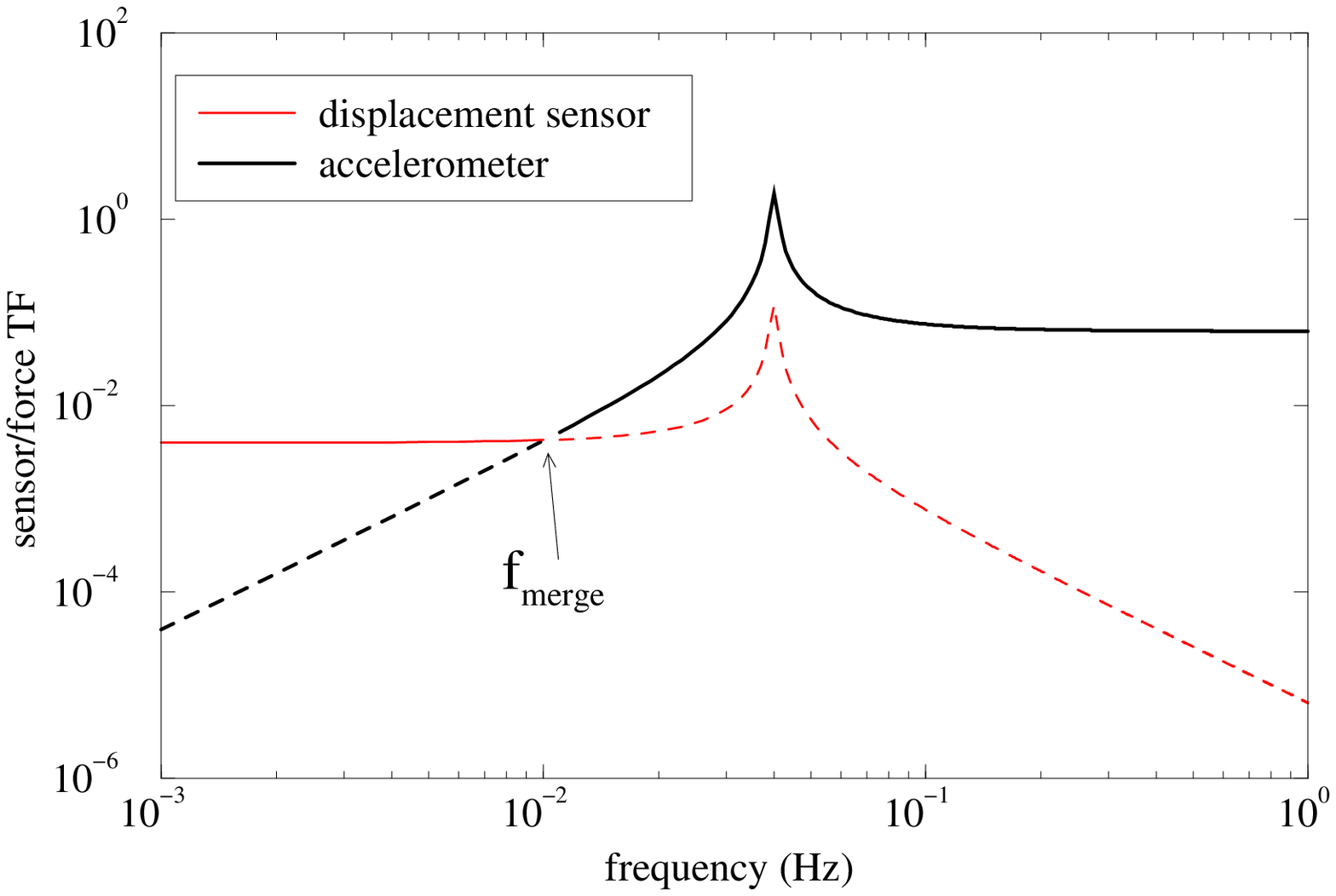}
\caption{\footnotesize {\it Merging} of displacement and
acceleration sensors (simulation for a simple pendulum).}
\label{merge}
\end{minipage}
\end{center}\end{figure}

\begin{figure}\begin{center}
\begin{minipage}{7cm}
\includegraphics[bb=92 43 737 490,clip=true,width=7cm]{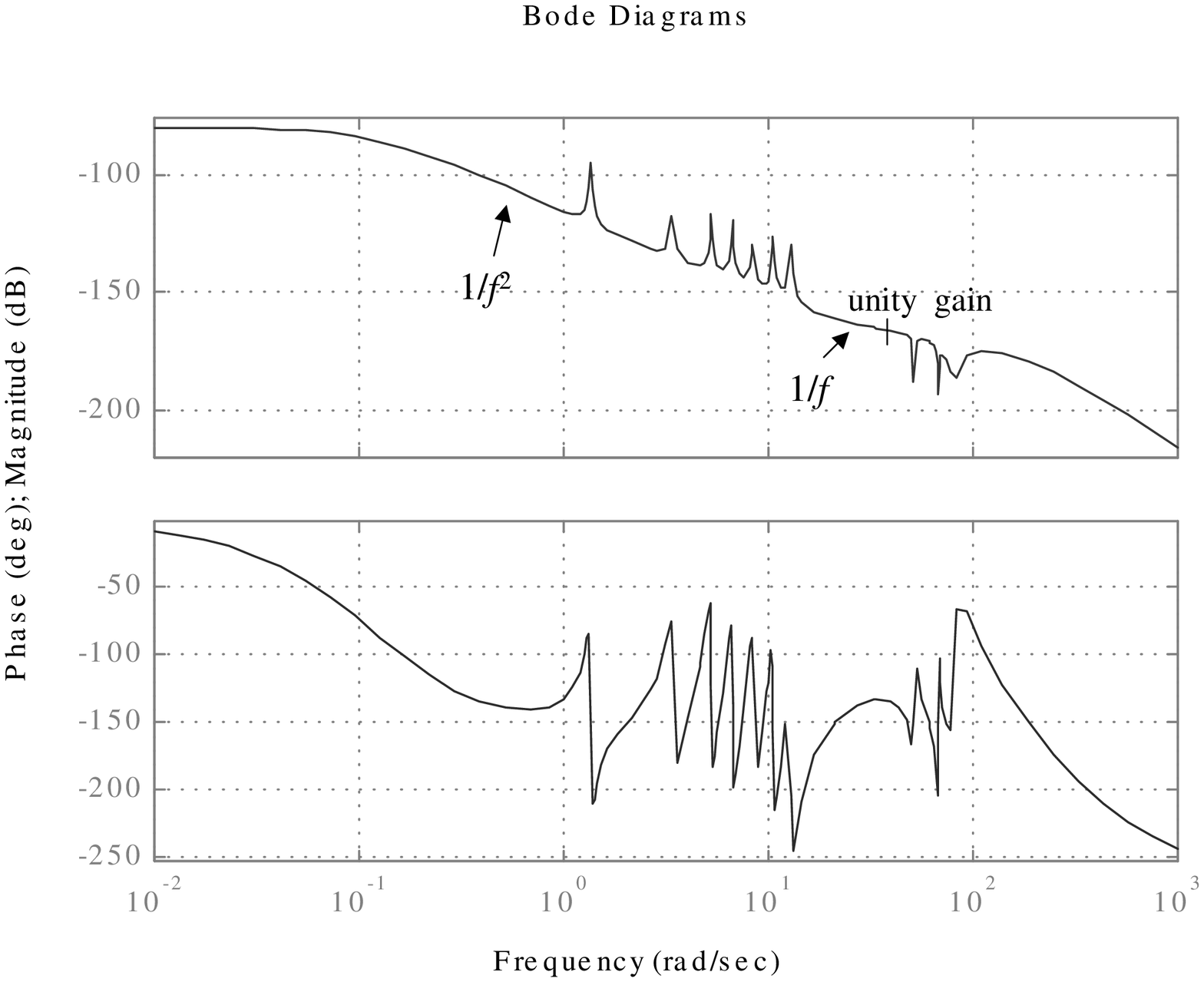}
\end{minipage}\hfill
\begin{minipage}{7cm}\includegraphics[width=7cm]{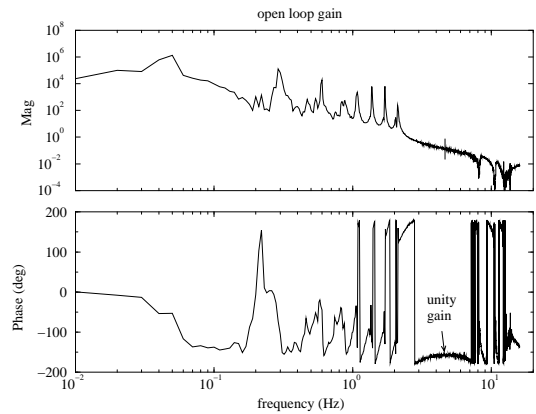}
\end{minipage}
\caption{\footnotesize LEFT: Digital filter used for the inertial
damping of a translation mode ($X$). The filter slope is $f^{-2}$
in the range 10 mHz$<f<$3 Hz, $f^{-1}$ for $f>$3 Hz. The unity
gain is at 4 Hz. The peaks in the digital filter are necessary to
compensate the dips in the mechanical transfer function (see the
transfer function of the $X$ mode in fig. \ref{xth}). RIGHT: open
loop gain function (measured). The phase margin at the unity gain
frequency is about $25^\circ$.} \label{filter}
\end{center}\end{figure}

 The object to control is a MIMO (multiple in-multiple out)
system: each sensor (accelerometer/LVDT) is sensitive to the 3
modes (X,Y,$\Theta$) of the IP and each actuator excites all the
modes. To simplify the control strategy the sensors outputs and
the actuators currents are digitally recombined to obtain
independent SISO (single in-single out) systems (fig.
\ref{decomp}): the system is described in the normal modes
coordinates (for a description of the diagonalization procedure
see \cite{nota,tesi}). Each normal mode is associated to a so
called {\it virtual sensor} (sensitive to that mode and ``blind"
to the others) and to a {\it virtual actuator} (acting on one mode
only, leaving the others undisturbed). In this way one is able to
implement independent feedback loops on each d.o.f., greatly
simplifying the control strategy. Fig. \ref{xth} shows the output
of the virtual accelerometers $X$ and $\Theta$. In the $X$ plot,
the 40 mHz resonance of the IP translation mode and all the modes
of the SA chain are visible (as pole/zero structures). In the
$\Theta$ plot, only the rotation mode of the IP is visible. The
two plots show that different feedback strategies have to be
implemented on the different d.o.f..  The basic idea of inertial damping is to use the
accelerometer signal to build up the feedback force. Actually, if
the control band is to be extended down to DC, a position signal
is necessary. Our solution was a {\em merging} of the two sensors:
the virtual LVDT and accelerometer signals are combined in such a
way that the LVDT signal ($l(s)$) dominates below a chosen cross
frequency $f_{\rm merge}$ while the accelerometer signal ($a(s)$)
dominates above it (see fig. \ref{merge} and ref. \cite{jila}).
The feedback force has the form\footnote{Actually, the LVDT signal
$l(s)$ is properly filtered in order to preserve the feedback
stability at the cross frequency and in order to reduce the amount
of reinjected noise at $f>f_{\rm merge}$.}: \begin{equation}
f_{\rm fb}=G(s)\left[a(s)+\epsilon l(s)\right] \end{equation}
where $G(s)$ is the digital filter transfer function (see fig.
\ref{filter}) and $\epsilon$ is the parameter whose value
determines $f_{\rm merge}$. We have chosen $f_{\rm merge}\sim 10$
mHz (corresponding to $\epsilon\sim 5\cdot 10^{-3}$). This
approach allows to stabilize the system with respect to low
frequency drifts at the cost of reinjecting a fraction $\epsilon$
of the seismic noise via the feedback.

\section{Inertial control performance}

\begin{figure}\begin{center}
\begin{minipage}{7cm}
    \includegraphics[width=7cm]{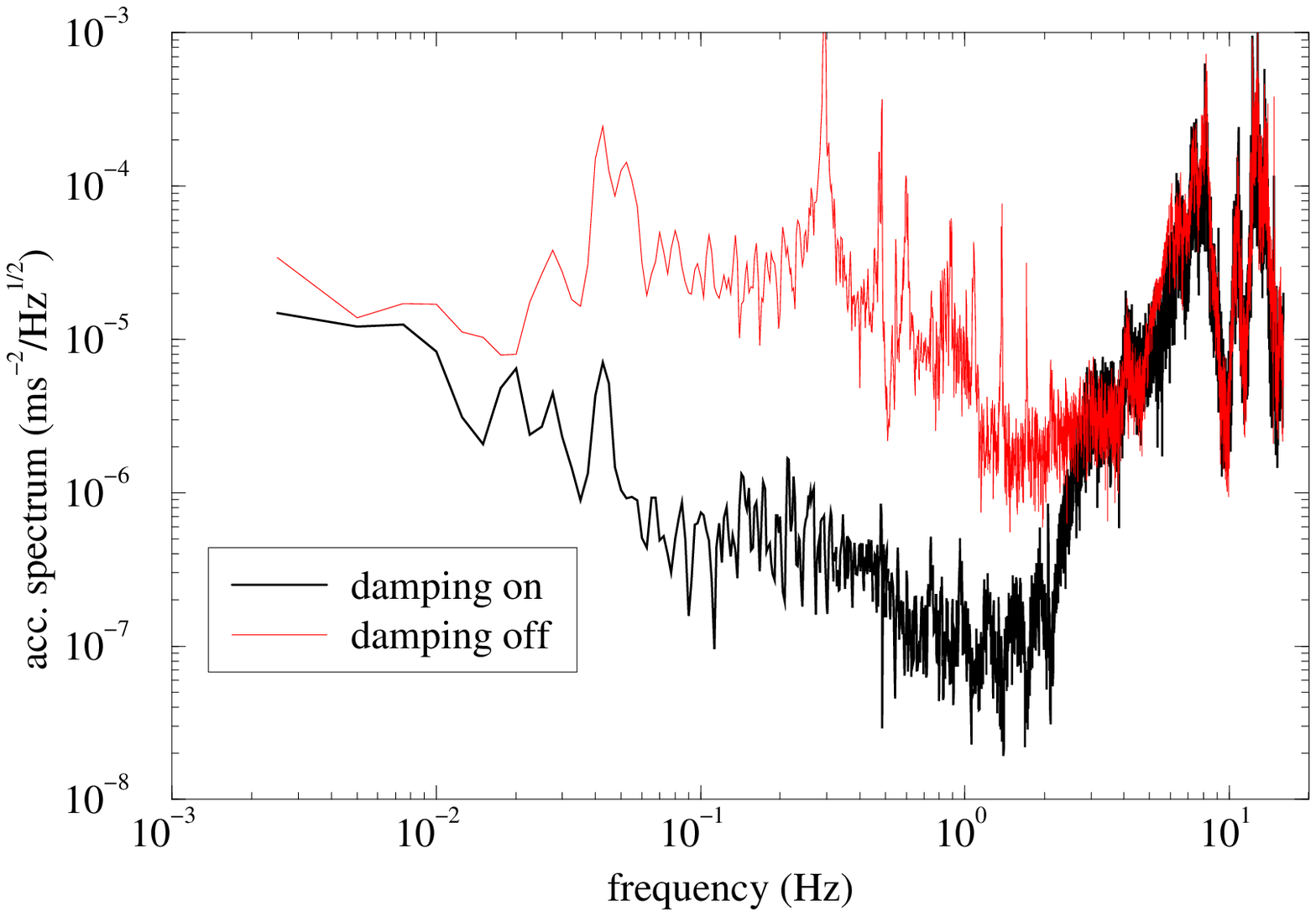}
\end{minipage}\hfill
\begin{minipage}{7cm}
 \includegraphics[width=7cm]{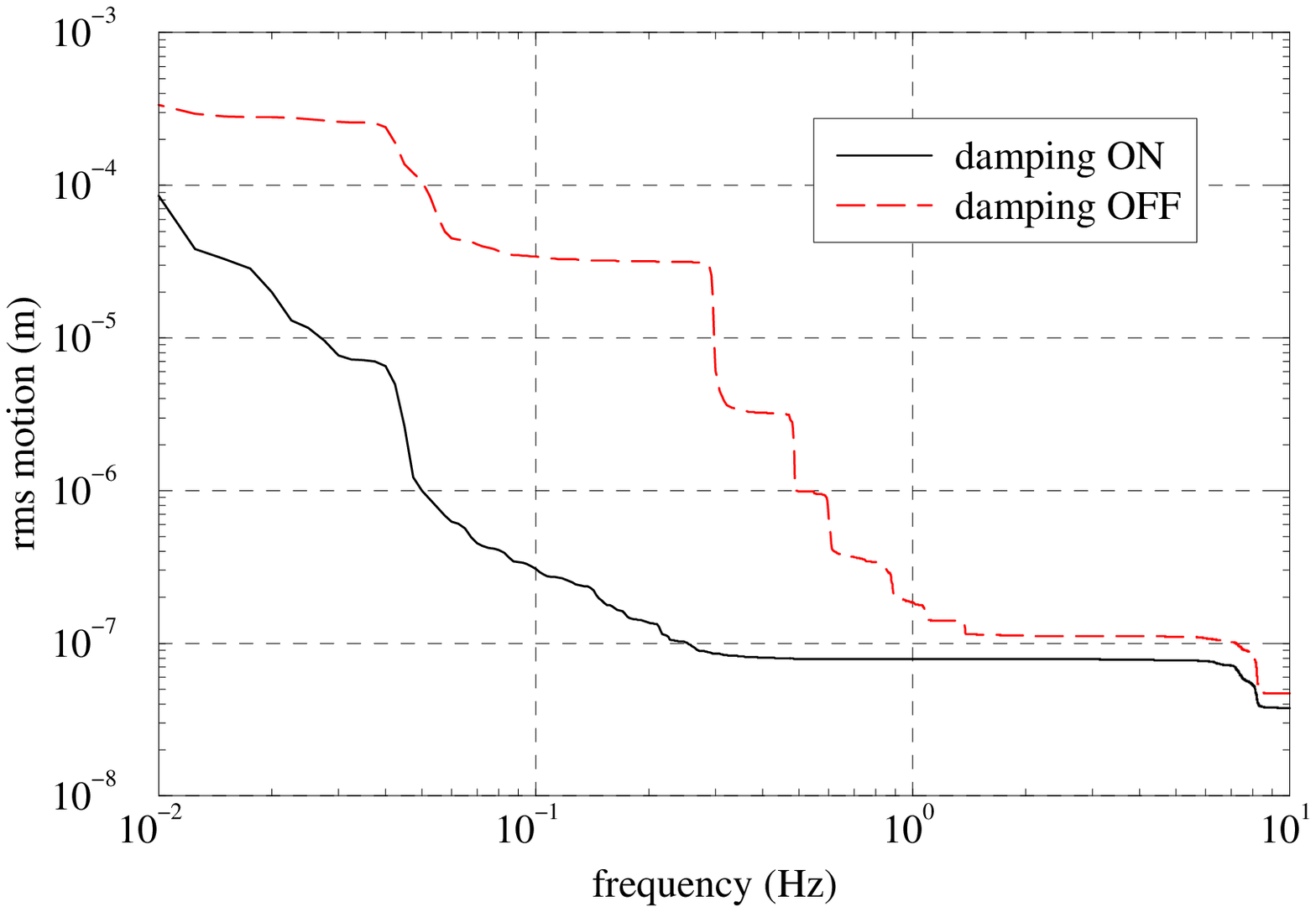}
\end{minipage}
\caption{\footnotesize Performance of the inertial control
($X,Y,\Theta$ loops closed) of the superattuenuator, measured on
the top of the IP: the left plot shows the acceleration spectral
density as measured by the {\it virtual} accelerometer $X$
(translation). The right plot shows the effect of the feedback on
the RMS residual motion of the IP as a function of the frequency.
} \label{results} \end{center}\end{figure}

The result of the inertial control (on 3 d.o.f.) is shown in
figure \ref{results}. The measurement has been performed in air.
The noise on the top of the IP is reduced over a wide band (10 mHz
- 4 Hz). A gain $>1000$ is obtained at the main SA resonance (0.3
Hz). The RMS motion of the IP (calculated as $x_{\rm
RMS}(f)=\sqrt{\int_f^\infty \tilde{x}^2(\nu){\rm d}\nu}$) in 10
sec. is reduced from 30 to 0.3 $\mu$m. The closed loop floor noise
corresponds to the fraction of seismic noise reinjected by using
the position sensors for the DC control and can, in principle, be
reduced by a steeper low pass filtering of the LVDT signal at
$f>f_{\rm merge}$ and by lowering $f_{\rm merge}$: both this
solution have drawbacks and need a careful implementation.

Preliminary measurements of the displacement of the mirror with
respect to ground have been performed in air, using an LVDT
position sensor. The residual RMS mirror motion in 10 sec.
is\footnote{This number has been obtained with a feedback design
less {\it aggressive} then the one of fig. \ref{filter}: the gain
raised as $1/f$, the cross frequency was 30 mHz and no
compensation of the dips was needed.}: \begin{equation} x_{\rm
RMS}(0.1\,{\rm Hz})\leq 3\;\mu{\rm m}. \end{equation} When the
damping is on such a measurement can provide only an upper bound
because the LVDT output is dominated by the seismic motion of the
ground.

\section{Further developments}

Several ways of improving the inertial damping performance have
been identified:
\begin{itemize}
\item a steeper low pass filtering of the LVDT output above $f_{\rm
merge}$ may reduce the amount of reinjected seismic noise. In
doing this one has to be careful to preserve proper phase
difference between the LVDT and accelerometer signals;
\item the lower $f_{\rm merge}$ the smaller the amount of
reinjected noise. Lowering $f_{\rm merge}$ is difficult due to the
mechanical tolerance on the parallelism of the IP legs: if the
legs are not perfectly parallel, the top table tilts slightly as
it translates. Therefore, the accelerometer signal is dominated by
the tilt below 15-20 mHz, making thus impossible to use the
accelerometers at very low frequencies. A technique for
subtracting the effect of the tilt (using the information provided
by the displacement sensors) has been defined and used to obtain
the results here described \cite{damp}. Cancelling the tilt effect
down to $\sim 5$ mHz makes us able to use the accelerometers down
to 10 mHz \cite{damp}. Stricter requirements on the IP legs
machining and improvements in the tilt subtraction technique may
allow a lower cross frequency.
\end{itemize}

\end{document}